\documentclass[preprint,aps]{revtex4}
\usepackage[dvips]{epsfig}
\begin{document}

\title{Bandwidth control of forbidden transmission gaps in compound structures with subwavelength slits}
\author{Diana C. Skigin \footnote{email: dcs@df.uba.ar}}
\altaffiliation{Also at Consejo Nacional de Investigaciones Cient\'{\i}ficas y T\'ecnicas (CONICET)}
\affiliation{Grupo de Electromagnetismo Aplicado,
Departamento de F\'{\i}sica,
Facultad de Ciencias Exactas y Naturales, 
Universidad de Buenos Aires, 
Ciudad Universitaria, Pabell\'{o}n I, 
C1428EHA Buenos Aires, Argentina}
\author{Hung Loui \footnote{email: hloui@sandia.gov}}
\affiliation{Sandia National Laboratories, P.O. Box 5800,
Albuquerque, NM 87185-1330, USA}
\author{Zoya Popovic \footnote{email: zoya@colorado.edu}}
\affiliation{Department of Electrical and Computer Engineering,
University of Colorado, Boulder, CO 80309-0425, USA}
\author{Edward F. Kuester \footnote{email: kuester@schof.colorado.edu}}
\affiliation{Department of Electrical and Computer Engineering,
University of Colorado, Boulder, CO 80309-0425, USA}

\date{\today}


\begin{abstract}

Phase resonances in transmission compound structures with
subwavelength slits produce sharp dips in the transmission response. 
For all equal slits, the wavelengths of these sharp transmission minima 
can be varied by changing the width or the length of all the slits.
In this paper we show that the width of the
dip, i.e., the frequency range of minimum transmittance, can be controlled
by making at least one slit different from the rest within a compound unit cell.
In particular, we investigate the effect that a change in the dielectric filling, 
or in the length of a single slit produces in the
transmission response. We also analyze the scan angle behavior of these 
structures by means of band diagrams, and compare them with previous
results for all-equal slit structures.

\end{abstract}

\maketitle

\section{Introduction}

The possibility of designing the transmission response of periodic structures with subwavelength slits has attracted the interest of the
scientific community since these systems show promise for applications
such as the characterization of attosecond pulses \cite{apb}, integrated polarizers \cite{jvs}, optical data storage and external storage media. 
Also, a great effort has been devoted to the modeling of scattering by
thick metallic plates with 1D \cite{Porto,Garcia-Vidal,Xie2} and 2D \cite{Popov} periodic perforations since the experimental demonstration of enhanced transmission by Ebbesen {\em et al.} \cite{Ebbesen,Ghaemi}.
Most of the studies carried out in connection with this phenomenon
consider simple structures, i.e., periodic structures 
in which a unit cell is defined by a single hole or slit. 

Recent simulations reported in
\cite{letter1_rd,letter2_rd,diana16} suggest that compound gratings (periodic structures comprising several slits in each unit-cell) can exhibit sharp transmission nulls when illuminated by $p$-polarized plane
waves of particular frequencies. These sharp dips correspond to the 
excitation of phase resonances in the structure, which arise from a 
particular phase and amplitude distribution of the magnetic field in adjacent
cavities of a unit cell, and are characterized by a field enhancement within
the slits. The behaviour of phase resonances in compound structures with subwavelength slits under oblique illumination has been studied
recently \cite{diana16}, where the authors show that new phase modes, and therefore new transmission dips, that were not allowed for normal illumination due to symmetry reasons, appear for non-normal incidence.

Phase resonances were first reported in structures comprising a finite number of cavities on a perfect conductor \cite{Veremey3,dv1}, and later observed in compound reflection gratings under normal \cite{sad1,sad2,sad3} and oblique \cite{rsad1} illumination. Le Perchec {\em et al.} analyzed the excitation of these resonances in a two-slit system \cite{LePerchec}. 
The additional parameters in the complex structure should allow for more degrees of freedom in the design of components such as nonlinear devices and frequency selective surfaces. Besides, for a selected
region of the electromagnetic spectrum, the behavior of these 
complex structures can be compared to the effect produced by a defect in 
a photonic crystal (PC) \cite{Joannopoulos}. Whereas a defect in a PC 
allows transmission within a forbidden band, a phase resonance in a compound transmission grating inhibits transmission in a highly transmitting region \cite{diana16}.  

Recently, experimental demonstration of phase resonances in infinite compound gratings has been reported 
\cite{Hibbins-fase}. In this paper, the authors confirm earlier predictions made by Skigin and Depine in 
\cite{letter1_rd} and show that a 
structure comprising three slits within each period, illuminated at normal incidence, exhibits a dip in the 
transmission response within each Fabry-Perot resonance peak.

The theoretical approach used in this paper to model the 1D compound transmission grating
is based on the Generalized Scattering Matrix (GSM) formulation. It basically consists
in expanding the fields in each region in its own eigenmodes and then matching them imposing the boundary conditions.
Oblique excitations of both thick and thin simple periodic frequency selective surfaces with arbitrary aperture cross-sections have been successfully analyzed with this approach \cite{GSM1,GSM2,GSM3}, which has been recently extended to deal with periodic structures comprising complex unit cells \cite{AP_Jacques,Jacques_thesis}. One advantage of the Extended Generalized Scattering Matrix (EGSM) formulation is that the individual GSM matrices describing any single discontinuity can be calculated separately and often {\em a priori} in a coordinate system best suited for an analytical solution. The assembly of GSMs describing a multitude of transverse discontinuities (i.e. multiple perforations per unit cell) forms a coordinate invariant EGSM that are easily cascaded via standard operators for problems having arbitrary longitudinal complexity. This abstraction alleviates tedious coordinate notation and facilitates fast parallel computation of electromagnetic scattering from highly complex unit cells.

The purpose of this paper is to show the design flexibility that compound
gratings give for getting particular responses. We investigate the effect
produced by a change in the dielectric constant and in the length of a 
single slit within each period. We also analyze the dependence
of the response on the angle of incidence. 

In Sec. II we summarize the extended generalized scattering matrix (EGSM) formulation used to solve the diffraction problem of a $p$-polarized
plane wave by a compound structure. In Sec. III we show numerical examples that illustrate the possibilities that these structures provide, by
varying the dielectric filling as well as the length of
the slits. We also show band diagrams that permit to visualize the angular
dependence of the resonant frequencies and the bandwidth of the gaps, and compare them with those
obtained for all-equal slits structures.
Concluding remarks are given in Sec. IV.

\section{The Extended Generalized Scattering Matrix formulation}

The EGSM is an extension of the Mode Matching-Generalized Scattering Matrix (MMGSM)
method that allows computationally efficient solutions to a larger class of 
scattering problems. The method is capable of efficiently analyzing the reflection and transmission 
properties of TE- and TM- polarized plane waves incident at arbitrary angles onto 
thin or thick arbitrarily shaped frequency selective surfaces, gratings with generalized profiles,
metal structures with multiple layers of dielectric fillings and/or coatings. The geometry of the
scatterer can be periodic or aperiodic, transmitting and/or reflecting.

In the present case, we consider an infinitely periodic one-dimensional compound structure: 
each period comprises several perfectly conducting wires and three equal-width slits between wires are
formed. Each slit can be filled with a different dielectric, as shown in Fig. 1. 
In what follows we summarize the fundamentals of the method. Further details are found in \cite{AP_Jacques,Jacques_thesis}.

The structure is divided longitudinally into slices or waveguide sections. These sections and the interfaces between them are 
represented by their equivalent EGSM $S_G$ and $S_T$, respectively, 
where the subscript denotes the number of a transition ($T$) or guiding section ($G$). 
To obtain the required transition and guided-section EGSMs, the fields in each slit are represented by modal expansions
and appropriate boundary conditions are enforced at each transition: continuity of tangential electric fields in the open 
regions, null electric field on the perfect conductor, and continuity of tangential magnetic field. The equations 
resulting from these conditions are projected in convenient bases, which yields a matrix equation with the modal amplitudes towards and away from the transition as unknowns. Once all the guided sections and transition matrices are found,
they are cascaded to find a relationship between the incident plane wave and the reflected and 
transmitted amplitudes \cite{AP_Jacques,Jacques_thesis}.

In the case of a uniformly filled waveguide with arbitrary longitudinal profile, where reflections from the gradual 
bends are negligible, it is sufficient to consider it as a straight section of the same waveguide having the length 
of the bent waveguide \cite{Lewin,Katsenelenbaum}. This is the case of the examples shown in Fig. 5. 

\section{Results}

The results shown illustrate the evolution of the phase resonances in a perfectly conducting compound thick-metal grating comprising 
three rectangular slits filled with different dielectrics. In the examples we consider symmetric structures in which the central slit can be different from the external ones in its dielectric filling (see Fig. 1) and eventually in its length (see inset 
of Fig. 5). A $p$-polarized plane wave impinges on the structure in all the examples shown.

In Fig. 2 we show a sequence of curves when varying the dielectric constant of the central slit filling ($\varepsilon_c$) while keeping all the rest of the parameters fixed: $c/d=a/d=0.08$, $h/d=1.14$, $\varepsilon=1$, $\theta_0=0^\circ$. We plot the zero-order transmittance as a function of the incident wavelength for six cases: $\varepsilon_c$ = 1, 1.2, 1.4, 1.6, 1.8, 2. The lower 
curve corresponds to identical slits, and the phase resonance appears as a sharp dip within the waveguide resonant peak of
the slits. Phase resonances are produced in compound gratings due to the possibility of having different configurations
of the field phase distribution within the different slits comprising each period of the structure \cite{letter1_rd}. In particular, for three
slits within the period, only one resonance is expected for normal incidence: the (+ - +) mode, i.e., the mode in which
the phases of adjacent slits are reversed. The number of possible phase configurations is finite and depends on the number of slits. A mathematical criterion governing the resonances using the EGSM formalism is given in \cite{AP_Jacques,Jacques_thesis}.

In general, two requirements are needed for a phase resonance to occur: i) at least one 
non-zero phase difference is found between the field phases in adjacent slits ---what is not allowed 
in simple periodic structures--- and ii) a particular distribution of the field amplitude,
naturally generated by the incidence conditions, is obtained \cite{dv1}. Phase resonances are
also characterized by a strong intensification of the field inside the slits or inside the cavities, in the case of
compound reflection gratings \cite{sad1}. In particular, when the phases in adjacent slits are opposite to each other, $\pi$ resonances can be excited \cite{dv1}. Regarding each slit as a waveguide,
these two conditions are usually fulfilled within the waveguide mode resonant wavelength.

It can be observed that as the permittivity of the central slit is increased, the width of each phase resonance dip 
also increases. 
This suggests that the bandwidth of the narrow gaps generated by means of phase resonance excitations
can be controlled. This characteristic makes this structure attractive from the point of view of several applications,
such as filters and polarizers. It can also be regarded as a complementary structure to the photonic crystal. A typical band diagram of a photonic crystal presents band gaps, i.e.,
frequency ranges that are not allowed to propagate inside the crystal and therefore, if the
structure is illuminated by a plane wave of a frequency within the gap, all the power is
reflected \cite{Joannopoulos}. However, if the perfect periodicity of the photonic crystal
is broken by a defect in the structure, allowed states arise within the gap, enabling transmission within the originally forbidden gap. In the compound gratings considered in this paper, the structures are essentially transmitting,
at least within the waveguide resonance peaks, as observed in Figure 2. Thus, when a
simple grating is illuminated by a plane wave of a wavelength within the waveguide resonance
peak, most of the power is transmitted through the structure. The addition of complexity to
the period in the form of a group of slits can be regarded as a defect in the 
perfect periodicity, and in this case one or more forbidden channels are formed within the allowed bands \cite{diana16}. 

As it was already investigated in \cite{diana16} for all-alike slits structures, when compound gratings are obliquely illuminated, new possiblities of phase configurations within the slits open up, resulting in new phase resonances. For instance, for three equal slits we have the (+ - -) and the (+ + -) modes in addition to the one we had for normal
incidence. To visualize the effect of changing the dielectric filling of the central slit in the response of the
structure under oblique illumination, we show in Fig. 3 the transmittance for the same structure considered in Fig. 2, but
for an oblique incidence: $\theta_0=30^\circ$. It can be observed that now there are two dips within each waveguide resonance 
maximum, i.e., around $\lambda/d \approx 1.5$ and $2.5$. For oblique incidence, the symmetry condition imposed by normal illumination is removed, and this allows new phase configurations inside the slits, which produce new dips in the transmission
response. The resonance dip at $\lambda/d$ slightly smaller than 2.5 for all equally filled slits (lower curve in Fig. 3(a))
shifts to larger wavelengths with decreasing quality factor ($Q$) for small variations of the central dielectric constant 
$\varepsilon_c$ (see the curve for $\varepsilon_c=1.1$). From then on, it collapses into a very high-$Q$ resonance for $\lambda/d$
slightly larger than 2.5, and stays fixed throughout all further increases of $\varepsilon_c$. The dip for $\lambda/d$ slightly
larger than 2.5 in the lower curve ($\varepsilon_c=\varepsilon$) shifts to larger wavelengths with decreasing $Q$ as $\varepsilon_c$
increases. This behaviour suggests that both dips behave differently for variations in $\varepsilon_c$. The persistence of the
high-$Q$ resonance dip (leftmost dip within the peak) for all dielectric variations suggests that it is fixed by geometry
and excitation, whereas the $Q$ and the position of the rightmost dip are dielectric filling dependent. Besides, the
transmittance peak to the left of the sharp null does not change much as $\varepsilon_c$ is varied, whereas the peak 
to the right of the rightmost null is certainly tunable by changing the relative dielectric constant of the
central slit filling.

To get more insight about the influence of changing the central slit filling when the structure is obliquely illuminated,
in Fig. 4 we show contour plots of the zero-order transmittance as a function
of $\alpha_0 d$ and $\omega d/c$, where $\alpha_0$ is the incident wave vector component along the direction of
periodicity and $\omega$ is the frequency of the incident wave. The lighter zones represent higher transmitted intensities. 
Fig. 4(a) corresponds to a compound structure with all-alike slits. It can be observed that there are mainly two transmission peaks, for $\omega d/c \approx$ 2.5 and 5. These maxima are scored by nulls, the number of which depends on the incidence
angles: for normal illumination there is one dip within each peak, whereas for larger angles two dips are found within each
peak. When the dielectric constant of the central slit filling is increased in steps of 0.2, we get the plots of Figs. 4(b) to 4(f) ($\varepsilon_c$ = 1.2, 1.4, 1.6, 1.8 and 2). It can be noticed that for normal incidence, the dip width increases with $\varepsilon_c$, as it was 
already observed in Fig. 2, and this trend is also maintained for the larger wavelenght dip at oblique incidence, as
observed in Fig. 3 for $\theta_0=30^\circ$. The results shown
in Fig. 4 suggest that the phase resonance mechanism could be exploited not only to design structures with a prescribed
number of resonances at specified wavelengths, but also to control their bandwidth by changing the dielectric filling.

The effect produced by changing the dielectric filling of the central slit in the transmitted response of the structure is
similar to what occurs when the length of the central slit is changed. 
The propagation constant of each mode of a slit bounded by perfectly conducting walls and filled with a material with
dielectric constant $\varepsilon_1$ is:
\begin{equation}
\beta_{1m}= \sqrt{\left(\frac{\omega}{C}\right)^2 \; \varepsilon_1 - \left(\frac{m \pi}{c}\right)^2}\;\;,  \label{beta1}
\end{equation}
where $C$ is the speed of light in vacuum and $m$ is an integer.
Then, for a given length of the slit $\rm{lz}_1$, the phase difference gained by the mode when going through the slit 
is $\beta_{1m}\, \rm{lz}_1$. If the slit is now filled with another dielectric with $\varepsilon_2= \alpha \,\varepsilon_1$, the new
propagation constant is 
\begin{equation}
\beta_{2m}= \sqrt{\left(\frac{\omega}{C}\right)^2 \,\alpha \,\varepsilon_1 - \left(\frac{m \pi}{c}\right)^2}\;\;, \label{beta2}
\end{equation}
and then the same phase difference would be obtained for another slit length $\rm{lz}_2$ such that 
\begin{equation}
\beta_{1m}\, \rm{lz}_1=\beta_{2m}\, \rm{lz}_2\;\;.    \label{betalz}
\end{equation}
In particular, if the slit has a subwavelength width, as in the present case, the only propagating
mode is the first mode, which corresponds to $m=0$ in (\ref{beta1}) and (\ref{beta2}). Then, condition (\ref{betalz}) yields 
$\rm{lz}_1= \sqrt{\alpha}\, \rm{lz}_2$.
Taking into account the above relationship between the length and the dielectric constant of the slit filling, 
it is to be expected that the response of a structure with a higher central permittivity should be similar to that of 
the structure comprising equally
filled slits, but with a longer central slit. The EGSM method used to solve the diffraction problem allows for the
treatment of bent waveguides with arbitrary longitudinal profile. Then, we increased the length of the central slit ($\rm{lz}$)
by bending it slightly, as shown in the inset of Fig. 5. The results in Fig. 5 show the evolution of the transmitted 
intensity as the central slit length is increased, when keeping $\varepsilon=\varepsilon_c=1$, for the same parameters considered in Fig. 2. It can be observed that the three curves ($\rm{lz}= h/d$, $\sqrt{1.2} h/d$ and $\sqrt{1.4} h/d$) for
$\varepsilon_c=1$ are very similar to those of Fig. 2(a), for $\varepsilon_c=1$, 1.2 and 1.4 and $\rm{lz}=h/d$. This result also shows that the resonance bandwidth can be
controlled not only by changing the dielectric filling, but also by varying the length of the slits.
 
\section{Conclusions}

We have investigated the response of compound transmission wire gratings with subwavelength slits under oblique
incidence. To solve the diffraction problem we used the EGSM method, which is very versatile. The ease of its formulation
significantly reduces the time and effort required for more advanced structures.
We focused in a structure comprising three slits in each period and analyzed the influence that a change
in the dielectric filling and in the length of the central slit has over the resonant dip. We have shown numerical 
results that evidence the possibility of controlling the bandwidth of phase resonance dips by properly choosing the
geometrical and constitutive parameters. The angular behavior of these structures was also analyzed by means of band 
diagrams, and the results obtained were compared with previous ones corresponding to all-equal slit structures.
The capability of exciting and controlling the bandwidth of phase resonances in compound gratings opens up new possibilities
for practical applications, such as polarization sensitive aperture shapes for field enhancement devices.

\section*{Acknowledgements}
 
D. Skigin gratefully acknowledges Prof. Ricardo Depine for discussions and support from Consejo Nacional de Investigaciones 
Cient\'{\i}ficas y T\'ecnicas (CONICET), Universidad de Buenos Aires (UBA, X150 and X2) 
and Agencia Nacional de Promoci\'on Cient\'{\i}fica y Tecnol\'ogica (ANPCYT-BID 1728/OC-AR-PICT 14099). 
Z. Popovic and H. Loui acknowledge support from a NSF International Supplement to a ITR Collaborative Research Grant No. CCR-0112591 and by the Department of Education Graduate Assistance in Areas of National Need (GAANN) Fellowship in Hybrid Signal Electronics (HYSE), award \#P200A040154. This research was also supported in part by an appointment to the Sandia National Laboratories Truman Fellowship in National Security Science and Engineering, sponsored by Sandia Corporation as Operator of Sandia National Laboratories under its U.S. Department of Energy Contract No. DEAC0494AL85000.

\newpage
\begin{figure}[h]
\begin{center}
\includegraphics[width=7in]{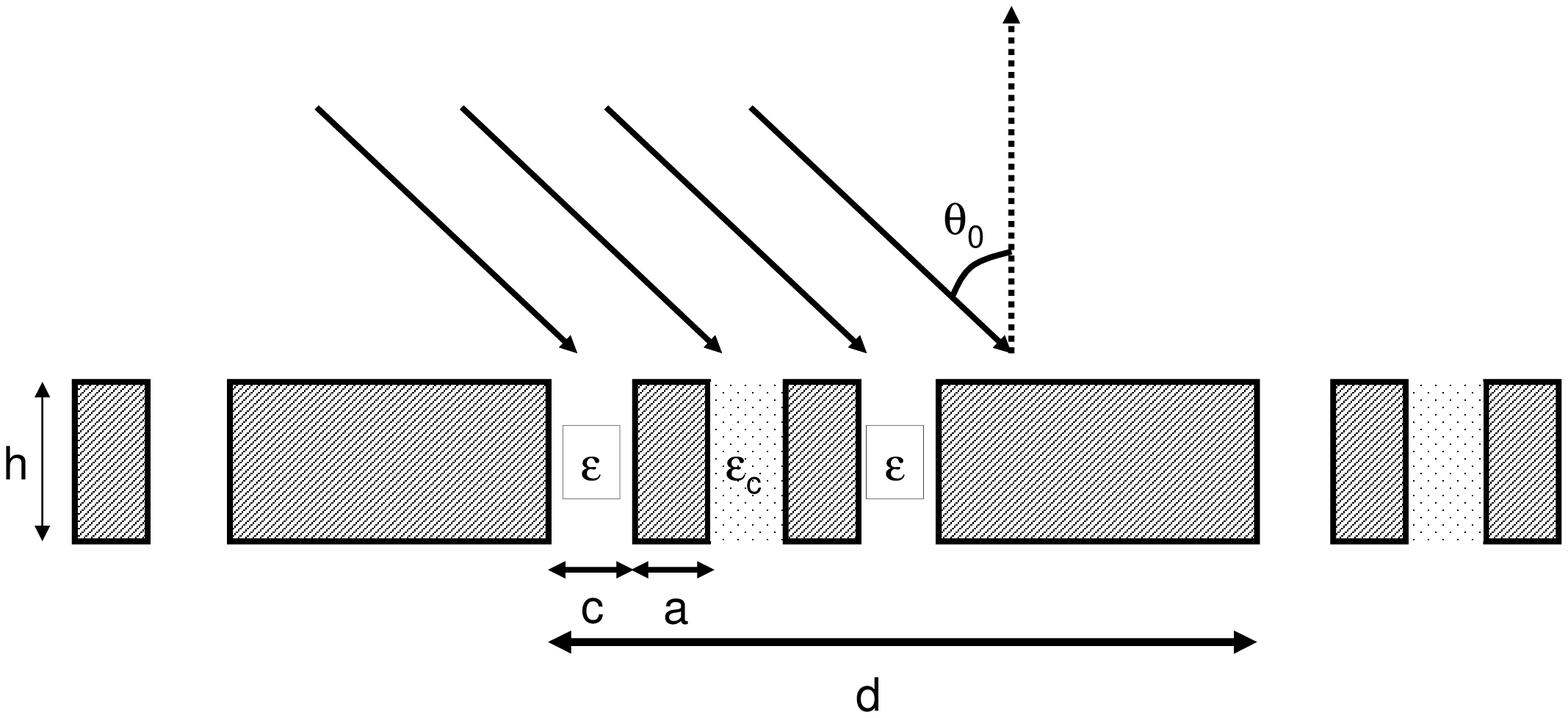}
\end{center}
\caption{Sketch of a three-slit compound grating. The relative permittivity of the material inside the middle slit differs from that in the edge slits. In this work, the slit length, $h$, and the slit widths, $c$ and $a$, are varied and the resulting response is found as a function of the incidence angle $\theta_0$. }
\end{figure}

\newpage
\begin{figure}[h]
\begin{center}
\includegraphics[width=25cm]{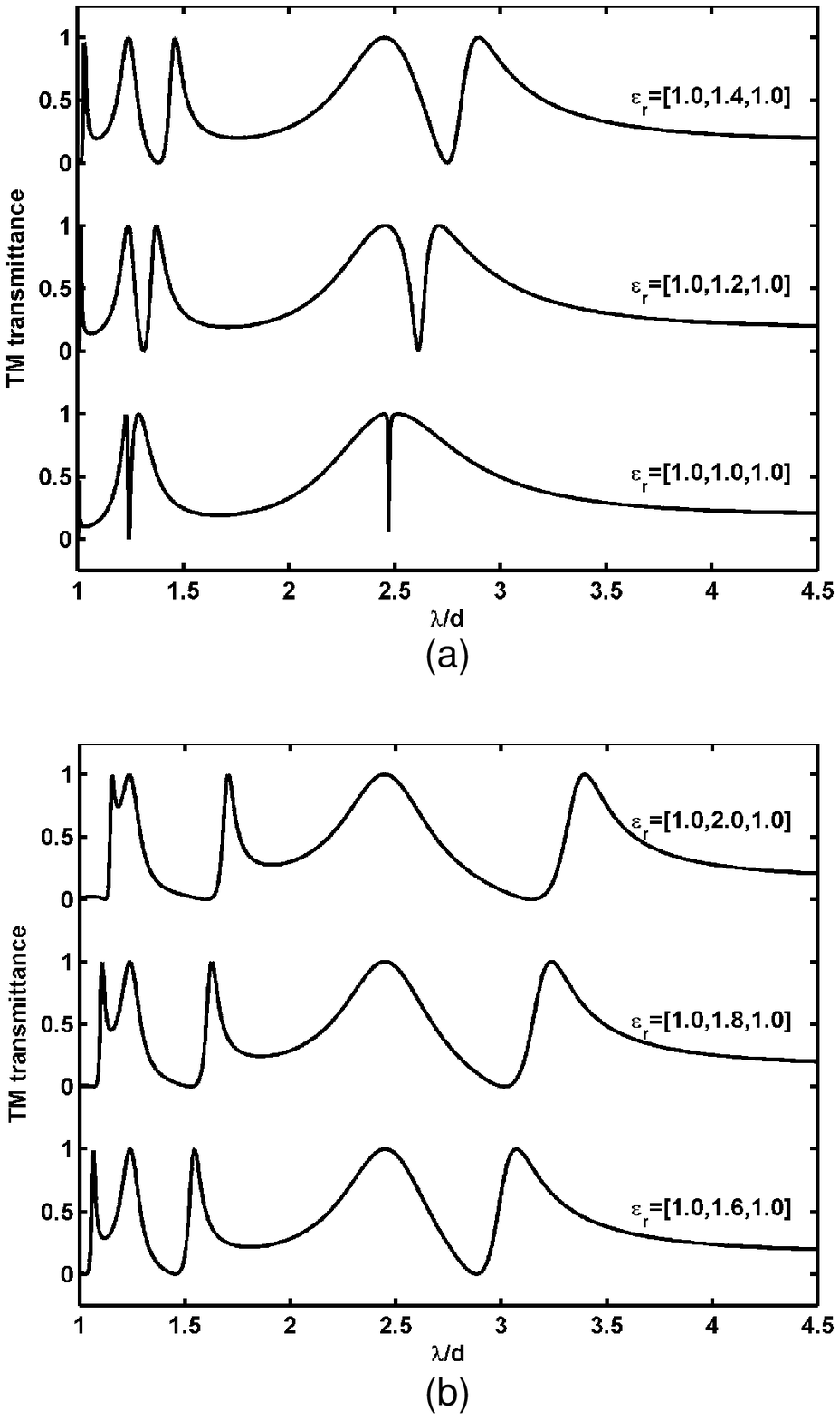}
\end{center}
\vspace{-0.5cm}
\caption{Zero-order transmittance as a function of the normalized incident wavelength for a structure with a three slit unit cell, 
incidence angle $\theta_0=0^\circ$, for six values of the dielectric constant of the central slit filling $\varepsilon_c$=1, 1.2, 1.4, 1.6, 1.8, 2. The rest of the parameters are: $c/d=a/d=0.08$, $h/d=1.14$ and $\varepsilon=1$.}
\end{figure}

\newpage
\begin{figure}[h]
\begin{center}
\includegraphics[width=15cm]{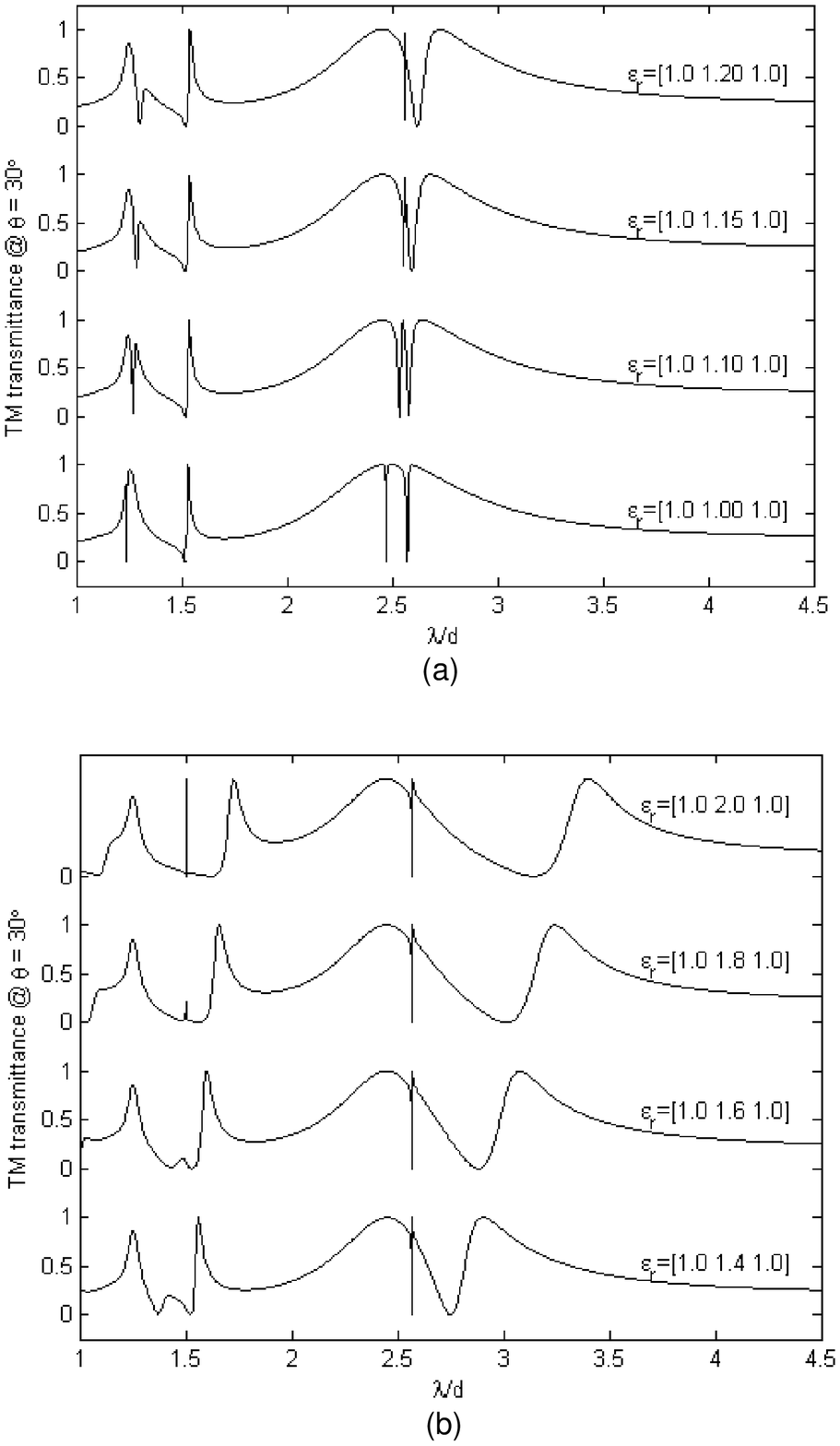}
\vspace{-3cm}
\end{center}
\caption{Zero-order transmittance as a function of the normalized incident wavelength for a structure with a three slit unit cell, 
incidence angle $\theta_0=30^\circ$, for several values of the dielectric constant of the central slit filling $\varepsilon_c$=1, 1.1, 1.15, 1.2, 1.4, 1.6, 1.8, 2. The rest of the parameters are: $c/d=a/d=0.08$, $h/d=1.14$ and $\varepsilon=1$, where $d$
is the period.} 
\end{figure}

\newpage
\begin{figure}[h]
\begin{center}
\includegraphics[width=15cm]{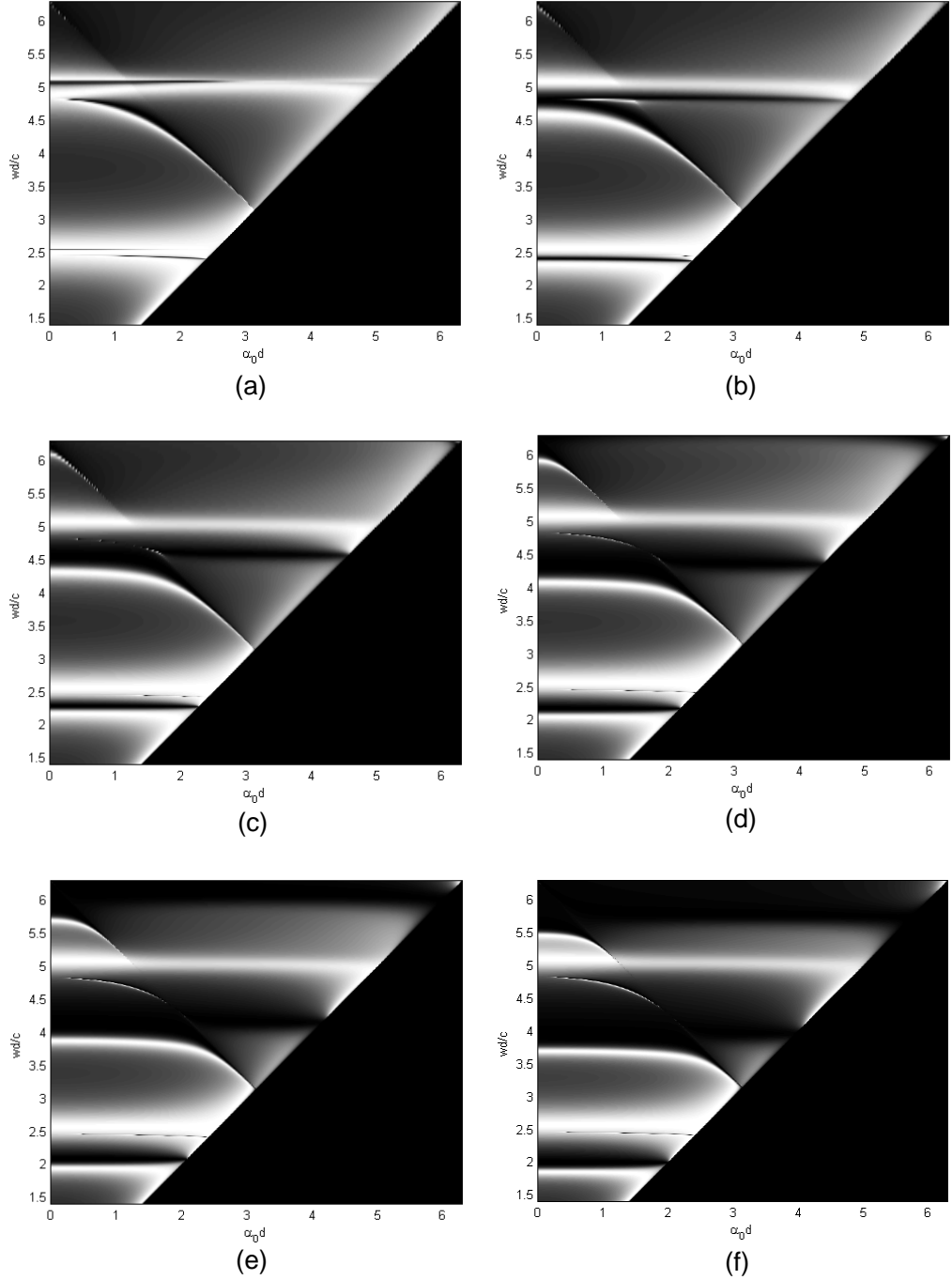}
\end{center}
\vspace{-2cm}
\caption{Contour plots of the zero-order transmitted intensity as a function of $\alpha_0\;d$, and 
$\omega\;d/c$, for the same cases considered in Fig. 2: (a) $\varepsilon_c=1$; (b) $\varepsilon_c=1.2$; (c) $\varepsilon_c=1.4$; 
(d) $\varepsilon_c=1.6$; (e) $\varepsilon_c=1.8$; (f) $\varepsilon_c=2$. Black represents null transmittance and white 
represents total (unity) transmittance.}
\end{figure}

\newpage
\begin{figure}[h]
\begin{center}
\includegraphics[width=15cm]{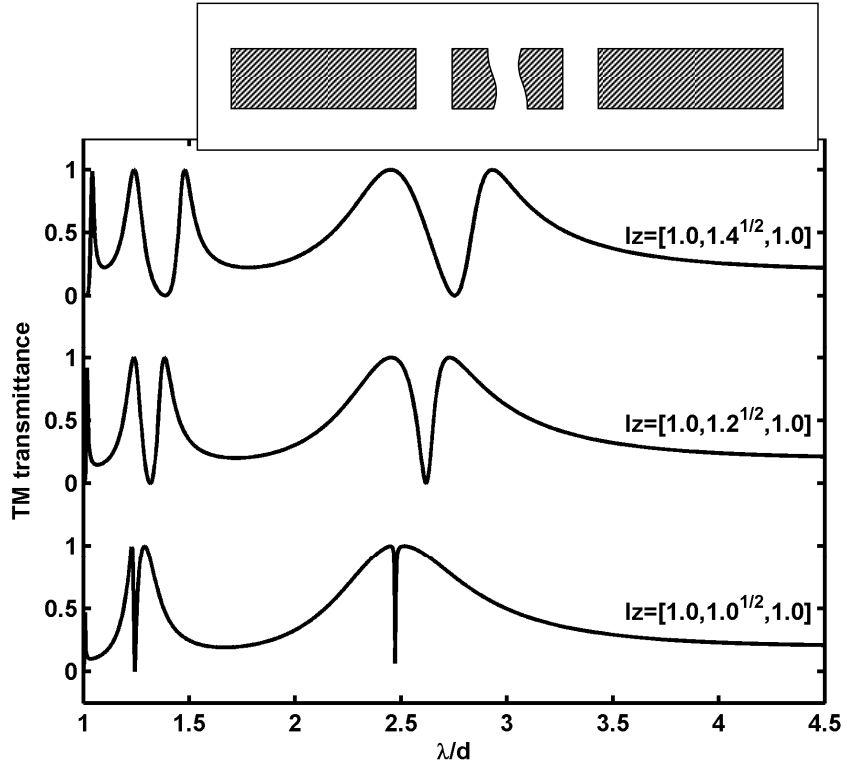}
\end{center}
\vspace{-4cm}
\caption{Zero-order transmittance as a function of the incident wavelength for a structure with a three slit unit cell, 
incidence angle $\theta_0=0^\circ$, for three values of the central slit length $\rm{lz}=h/d$, $\sqrt{1.2}h/d$, $\sqrt{1.4}h/d$. 
The rest of the parameters are: $c/d=a/d=0.08$, $h/d=1.14$ and $\varepsilon=\varepsilon_c=1$. The inset shows the scheme 
of a structure with a longer central slit.}
\end{figure}

\end{document}